\documentclass[preprint,aps]{revtex4}
\usepackage{graphicx}
\usepackage{epsfig}
\begin{document}
\newcommand{\ofig}[2]{
\begin{figure}[p]
\includegraphics[width=13cm]{#1}
\caption{#2}
\label{fig:#1}
\end{figure} }
\newcommand{\lt}{<}

\title{A one-to-two dimensional mapping using a partial fast Fourier transform}
\author{Stellan \"Ostlund}
 \email{Stellan.Ostlund@physics.gu.se}
\affiliation{
Department of Physics \\
University of Gothenburg\\
Gothenburg SE-41296 \\
Sweden
}

\date{\today}

\begin{abstract}
It will be shown how to map a simple one-dimensional tight binding
model with a cosine potential in one dimension exactly to a two dimensional
tight binding model with periodic boundary conditions with the presence
of a single flux quantum spread evenly on the torus. The mapping
is is achieved by a partial sequence of ``Fast Fourier
Transform'' (FFT) steps which if completed would be an exact Fourier
transform of the original model. Each step of the FFT recursively maps a
tight binding model into two decoupled sublattices of half the lattice 
length.
\end{abstract}

\maketitle

\section{\label{sec:intro} Motivation}
The following paper is dedicated to Nihat Berker and describes a
mathematical curiosity that I stumbled across a few years ago. It
brings together two ideas that were a daily topic of discussion
with Nihat during my graduate student days: duality and  renormalization.
The work also resulted from my admiration of the elegant recursive
algorithms that lie at the basis of the efficient solution of a
number of practical problems, most notably sorting and the Fourier
transform.

For a transitionally invariant problem, a Fourier transform usually
either solves or at least decouples the spatial degrees
of freedom so that the original problem becomes simple. When applied
directly on data of length $ N $ it uses a number of operations
that scale like $ N^2 $.  The ``Fast Fourier Transform'' (FFT)
dramatically reduces the number of arithmetic operations using a
clever recursive algorithm. An interesting question would be if
if a single FFT recursion could somehow be the basis of
a renormalization transformation that would partially decouple a
problem at each stage of recursion. Ultimately I was not able to
answer this question but in the process came up with a surprising
result; that repeated application of the FFT recursion can map a
problem from one to two dimensions.

\section{\label{sec:background} Background}
Before proceeding to the calculation this I make a
brief comment on the history of the FFT. In 1965 Cooley and Tukey
wrote a famous paper entitled {\it ``An Algorithm for the Machine
Calculation of Complex Fourier Series'' }. 
\footnote{ J.W. Cooley
and J.W. Tukey, {\em Math. Comp } Vol. {\bf 19}, no. 2, 297-301
(1965) } 
Cooley and Tukey cited I.J. Good \footnote{ I.J. Good,
{\it ``The Interaction Algorithm and Practical Fourier Analysis'' }
, Journal of the Royal Statistical Society, Series B,  Vol.{\bf 20 }, no. 2, 361-372, (1958).
} as having influenced their work. The impact of the ``Cooley-Tukey algorithm'' 
was enormous and their ``Fast Fourier Transform'' immediately became
indispensable tools for engineers and scientists as computers started
to become widely available at this time. These days the ``FFT''
algorithm is part of every numerical package and most of us don't
pay much attention to the algorithm.

Soon after the paper came out it was realized that the ``FFT'' algorithm had been repeatedly
rediscovered earlier. A detailed historical discussion on the subject is available by
Heideman, Johnson and Burrus.\footnote{ Heideman, M. T., D. H. Johnson, and
C. S. Burrus, {\it "Gauss and the history of the fast Fourier transform," },
Archive for History of Exact Sciences, Vol. {\bf 34} , 256-277 (1985).  } The ``FFT'' was
eventually traced back to C.F. Gauss. It appears
certain that he discovered the algorithm around 1807 but did not publish it. 
Given the impossibility of making significant numerical calculations at 
the time Gauss probably did not consider the result particularly important. 
The algorithm was eventually published posthumously in his collected
works. \footnote{ C. F. Gauss, {\it "Nachlass, Theoria Interpolationis Methodo
Nova Tractata,"} in Carl Friedrich Gauss Werke, Band 3, K\"oniglichen
Gesellschaft der Wissenschaften: G\"ottingen, pp. 265-330, 1866.}
Thus Gauss' original work was done approximately simultaneously
with Fourier's development of his famous trigonometric expansion.
In my opinion though, Cooley and Tukey deserve great credit for realizing
in 1965 the importance of an efficient numerical Fourier transform
and deriving the algorithm  at a time when it became useful.

The power of the FFT algorithm is that it dramatically reduces the
number of arithmetic operations required for a Fourier transform
of a large data set. For the most common ``radix-2'' algorithm the
number of data points is taken to be $ L = 2^n $. The number of
operations will be on the order of $ 2 n L $ rather than the $ L^2
$ number of operations that results from applying a simple matrix
operation. The FFT algorithm can also be implemented on arbitrary
length data where the number of arithmetic operations scales like 
$ L $ times the sum of prime factors of $ L $. When $ L $ is prime,
the FFT reduces to the ordinary matrix operation.

Without going into detail, I will now write down the key formulas
using radix-2 algorithm that shows how the FFT is based on recursion.
We begin with data $ x_m $ where $ 0 \le m \lt L  $ and $ L =
2^n $. We define the Fourier transform
\[
y_k = \frac{1}{\sqrt{L}} \sum_{m=0}^{L - 1 } e^{ - 2 \pi i j k / L } x_j .
\]
The sum can be split into two parts:
\begin{eqnarray*}
y_k & = & \frac{1}{\sqrt{2}} ( \sum_{m=0}^{L/2-1} x_{2m} e^{ \frac{ 2 \pi i }{ L/2} m k } + e^{ \frac{ 2 \pi i }{L} k } \sum_{m=0}^{L/2-1} x_{2 m + 1 } e^{ \frac{ 2 \pi i }{ L/2} m k } ) \\
 & = & e_k + e^{ \frac{ 2 \pi i }{L} k } o_k 
\end{eqnarray*}
where $ e_k $ can be seen to be the Fourier transform of the even indexed subset of 
$ x_k $ and $ o_k $ is the Fourier transform of the odd subset. 
This procedure can therefore be done recursively, splitting $ e_k $ and $ o_k $
successively into into odd and even subsets each with half as
many data points.

This notation is clumsy and does not lend itself to recursion. To
implement it systematically we must keep track of which order of
even and odd blocks we use to construct $ x_k $ as we apply the
algorithm. We use the label $ c_{j,k} $ keep track of our recursive
indexing. We begin by defining $ c_{j,1} = x_j $ to consist of the
original data. The recursion then consists of breaking the first
index $ j $ into odd and even blocks. Since I will shortly write down a
precise formula, I will first just the sketch the idea of constructing
the indexing. We lay the ``odd'' and ``even'' blocks next to one
another and put in a binary index ``0'' last if the block is even
and ``1'' if the block is odd. The original block is periodic in
length $ L $; each new block is periodic in length $ L/2 $ and the
number of blocks, say $ N $ in number is doubled to length $ 2 N$. 
Thus the original index $ (j,1) $ will be replaced by an index 
$ (j^\prime,k) $ where the ``block'' index $ k $ encodes odd and
even block and $ ( j^\prime ) $ describes the new order in the
subblock. Doing this recursively, the index $ k $ will gradually
become the ``frequency'' or ``momentum'' index. The recursion is
finished when the entire space index has been reduced to length one
and the momentum index has become length $ L $. \footnote{ The
block indexing naturally eventually generates the Fourier index
in bit reversed order, which usually requires a permutation operation
to be included in standard FFT algorithms. The reordering operation
does not complicate the present formulation.}

\newcommand{\comma}{ \, , \, }
Let us use the notation $ r:L $ to denote the integer
$ r $ modulo $ L $. We encode the algorithm as follows: 
\begin{equation}
c_{ r:L \comma  s:N} = \frac{1}{\sqrt{2}} \left( c_{r:L/2 \comma s: 2N} + e^{ 2 \pi i r / L } c_{r:L/2 \comma (s+N):2N } \right) 
\end{equation}
The inverse is given by 
\begin{equation}
c_{r:L, s:N} = \frac{1}{\sqrt{2}}\left\{
\begin{array}{cccccc} & ( c_{r:2L \comma s:N/2 } & + & c_{( r + L ):2L\comma s:N/2 } ) & \; if & \; s \lt N/2 \\ 
	 e^{ - \pi i r/L} & ( c_{r:2L \comma  s - N/2 :N/2 } & - & c_{( r + L ):2L \comma  s - N/2 :N/2 } ) \; & if & \; s \ge N/2 
\end{array} \right.
\end{equation}

It is straightforward to apply this operation to quantum mechanical
operators $ c_{ r\\:L, s\\:N} $ where $ c$ is simply the one particle
(destruction) operator on an $ L \times N $ lattice of "pseudo real
space" dimension $ L $ and "pseudo k-space" dimension $N$.
For the initial iteration $ L $ is the original
space dimension, and $ N = 1 $. Equivalently
we can simply
interpret $ c_{r}$ as the column vector being transformed and 
$ c^\dagger_{r} $ as a row vector.

Thus we begin with a periodic ``real space'' lattice of length $ 2^n $ to 
which we initially attach a bogus ``momentum space'' lattice
of length $1$ to begin the recursion with $ c_{j,1} $. At each FFT
step we halve the length of the real space and double the length
of the momentum space. Eventually when we are done applying the
recursion $ n $ times the ``real space'' lattice will have shrunk
to length one and the variables $ c_{1,k} $ that correspond to the
momentum space lattice will be the Fourier transform of the original
variables.

What if we stop this recursion ``half way''? Provided $ L $ is $ 2^{2m} $ 
where $ m $ is integer, we will obtain a two dimensional
periodic $ 2^m \times 2^m $ lattice index when we stop the procedure
after $ m $ recursions and thus will have obtained a mapping to a
``two-dimensional'' $ 2^m \times 2^m $ index $ c_{j,k} $ on a torus
from our original $ 2^{2m} $ periodic index.

Clearly this transformation is highly nonlocal since it mixes Fourier
components of indexes very far apart. If we start with an arbitrary
real space operator written in terms of the real space operators 
$ c_{r} $ and apply this prescription it is tempting to conclude that
we will always get a terrible mess during the recursion. This is
indeed true generically. However, we know that a transitionally
invariant operator is diagonalized by the Fourier transform, i.e.
written as a simple local operator in the fully transformed
basis. Thus eventually after recursion the momentum space blocks
must decouple and the intuition that the problem gets more and more
intractable under recursion is violated. Furthermore, 
a model which is invariant under Fourier transform should be symmetric 
in momentum and space coordinates half way through the recursion, so perhaps
such a model will retain a structure during the intermediate steps
of the recursion as well. It is precisely question we will explore
further by example.

In the rest of this communication, I will focus on the following model under
the ``partial'' FFT operation. We begin with
\begin{equation}
 h_{N\times 1} = - \sum_{r=0}^{N-1} ( c^{\dagger}_{r} c_{r+1:N} + cc ) - 2 \sum_{r=0}^{N-1} c^{\dagger}_{r} c_{r} \cos{2 \pi r / N } .
\end{equation}
This is in fact a special case of ``Harper's model'' where periodicity
is exactly  one relative to the length of the lattice.
\footnote{ P.G. Harper, {\em Single band motions of conduction electrons in a uniform magnetic field }, Proceedings of the Physical Society of London, Sec. A,
{\bf 68 }, 874-878 (1955)}
a point which I will discuss at the end of this paper.

We know that the first term transforms to the cosine under Fourier transform and vice versa. For
this choice of parameters the model is ``self dual'' i.e. transforms to itself under Fourier transform. 
It can therefore be expected to be symmetric in some way when partially transformed $ n $ times, where $ 2^{2n} = L $. 
For illustration, we begin with a $ 64 \times 1 $ lattice 
and apply the FFT step three times to transform this model numerically 
to an $ 8 \times 8 $ lattice. The result of applying 
this transformation is given by the following Hamiltonian
\begin{equation} 
h_{8\times 8} = - \sum_{r,\delta} t_{r,\delta} c^\dagger_{r} c_{r+\delta} .
\end{equation}
where $ \delta $ are only nearest neighbor lattice vectors,
where $ | t_{r,\delta} | = t $
and the phases of $ t $ correspond to a single flux evenly spread throughout the
entire lattice.\footnote{This is asserted as an unproved 
numerical observation.} 
The value of $ t_{r,r^\prime} = e^{ 2 \pi i n_{r,r^\prime} / 64 } $ 
where $ n_{r,r^\prime } $ is shown in Fig. \ref{fig:hcentered}.

\ofig{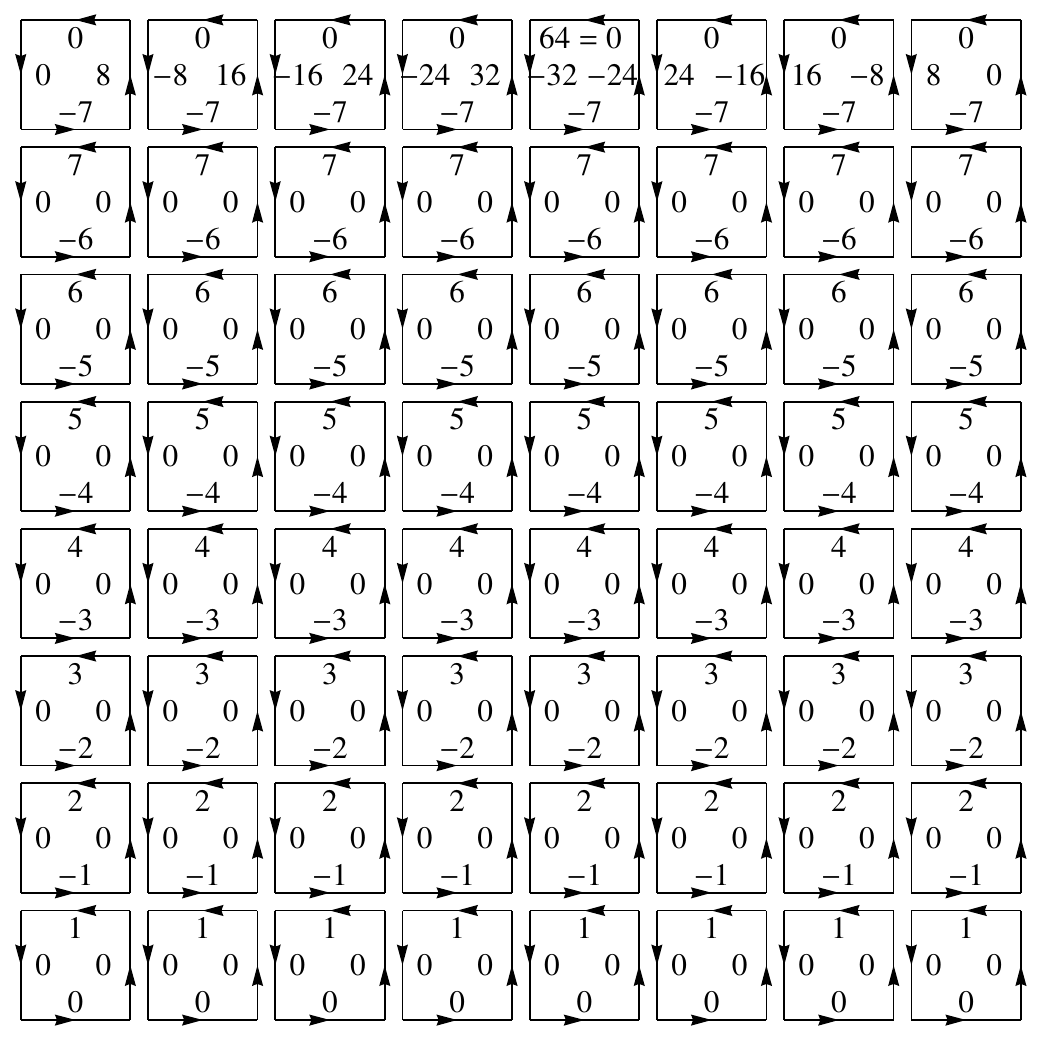}{Integers $ n_{r,r^\prime} $ where $ t_{r,r^\prime} = e^{ 2 \pi i n_{r,r^\prime} / 64 } $ are shown. } 
The vertical bonds with phase zero represent the mapping of the cosine potential,
whereas the horizontal bonds that carry the flux is the transformation
of the hopping. 

Let us look more carefully at the distribution of fluxes. With the
single exception of a plaquette in the top row, we note that the
sum of integers around each plaquette is one, so that the product
of phases around each plaquette is exactly $ e^{ 2 \pi i / 64 } $
which I interpret as a single flux quantum evenly spread on the
torus. Periodic boundary conditions are fulfilled and the left and
right boundary are mapped to each other, likewise for top and
bottom. The rule that the sum of phases around each
individual plaquette is the sum of phases around the entire boundary
is obtained form the fact that a plaquette on the top row has flux
$ -63 \approx 1 $. An equivalent phase choice of $ 2 \pi $ instead
of  zero  for the top link is equivalent to assigning a flux of
$ 2 \pi / 64 $ through each plaquette.

Although I do not present further results here, I assert that
similarly applying the FFT recursion a different number of times result
in rectangular lattices all with a total $ 2 \pi $ flux threading 
each periodic lattice.

It is interesting to compare the eigenstates of the tight binding model to
the corresponding states mapped by the partial FFT. The absolute square of 
the six lowest eigenstates $ \psi_{j} $ of of $ h_{256 \times 1 } $ is plotted 
in Fig.  \ref{fig:onedplots}.
\ofig{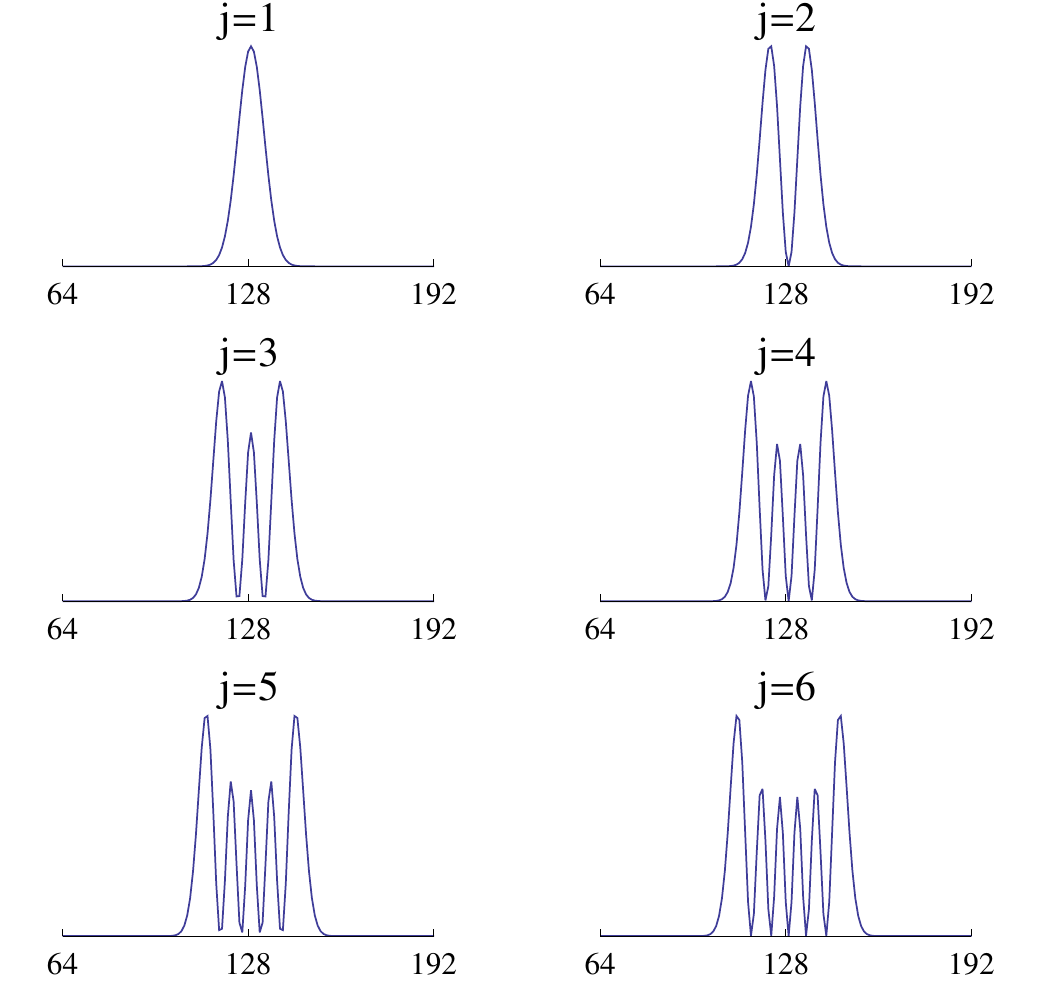}{Plot of $ |\psi_j|^2 $ for the six lowest energy eigenstates of $ h_{256\times 1 } $. }
These corresponding eigenstates of $ h_{16\time 16} $
is plotted in Fig. \ref{fig:vortexplots} and thus represents
the ``2-D'' mapping of the ``1-D'' states in Fig. \ref{fig:onedplots}.
\ofig{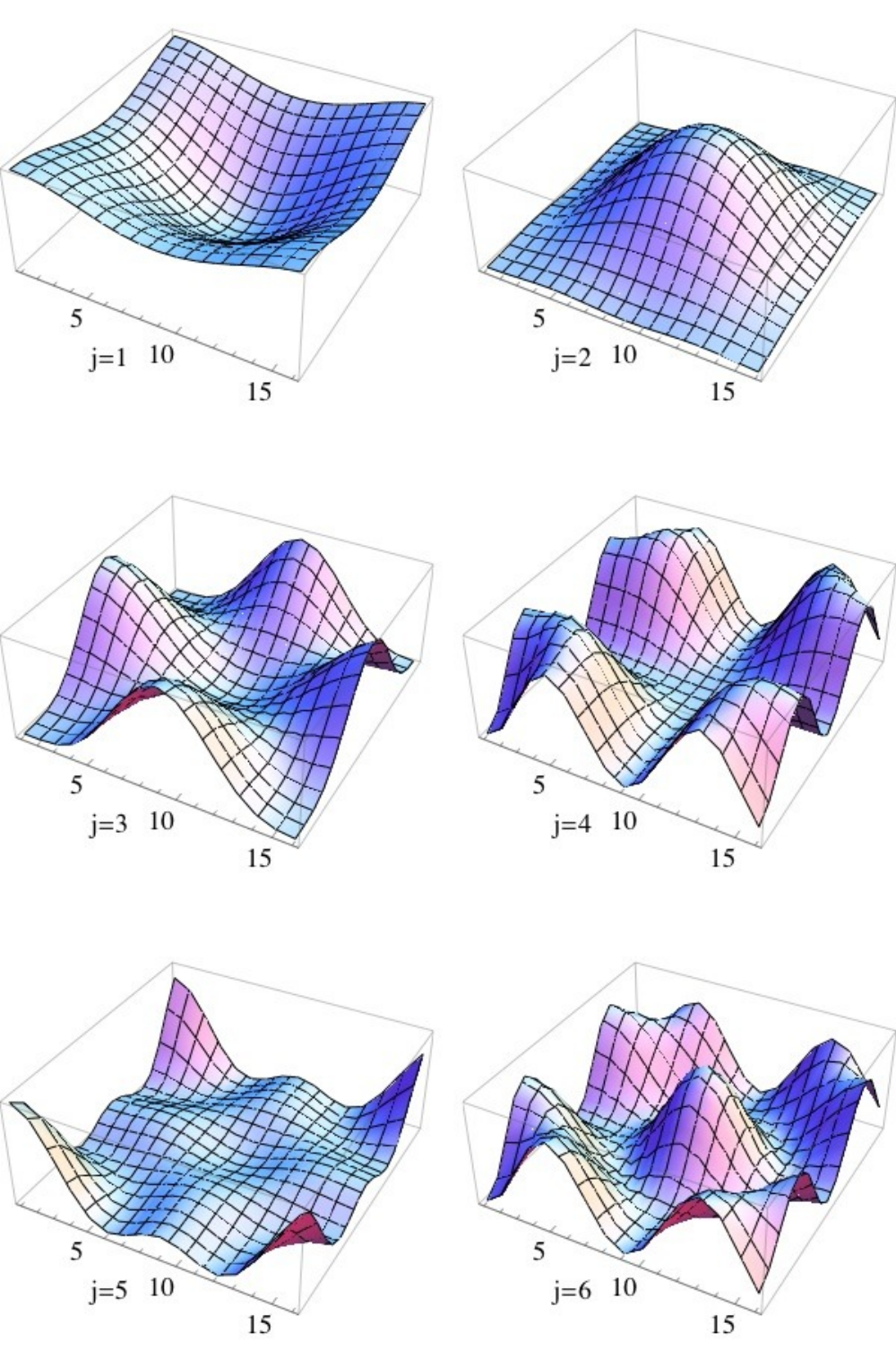}{Plot of $ |\psi_{j,k}|^2 $ for the six lowest energy eigenstates of $ h_{16\times 16} $ mapped to the two dimensional lattice }
The calculations are displayed here for values of $ N $ which are
not large; analogous results for larger values of $ N $ have been
computed.

\section*{Connection to Harper's equation} It is well known that
an infinite two-dimensional tight binding model in a constant
magnetic field maps to a one tight binding model with a periodic
or quasiperiodic potential.  In addition to P.G. Harper's paper
mentioned previously, there are several other early classic papers on 
listed in the references.\footnote{ R. Peierls, {\it Z. Phys.
80 } 763, (1933) ; M. Ya. Azbel, {\it Zh. Eksp. Teor. Fiz. }{\bf 44}, 980 (1963);
[Sov. Phys.—JETP 17, 665 (1963)] } 

In 1976, R. Hofstadter's performed a numerical analysis of the problem
and pointed out the  beautiful self-similar structure of the energy spectrum.
\footnote{R. Hofstadter, {\it Phys. Rev. }{\bf  B. 14} , 2239, (1976) }
A number of paper followed in the 80's that explored this
using renormalization group ideas.\footnote{M. Wilkinson, {\em J Phys A-Math Gen }, {\bf 20}(13), 4337-4354 (1987); S. Ostlund and R. Pandit, {\em Phys. Rev.}{\bf B 29 }, 1394 (1984)} The extent
to which these are pertinent to the present problem is addressed
in this section.

For a tight-binding model in the presence of a magnetic field with 
flux $ \alpha / 2 \pi $  per plaquette we can introduce a Landau gauge 
and write the Hamiltonian as
\begin{equation} \label{eq:harptwod}
 h  = - \sum_{r,\delta} t_{r,\delta} c^\dagger_{r} c_{r+\delta} 
\end{equation}
where $ t_{r, \pm \hat{y} } =  t $  and 
$ t_{r,\mp \hat{x} } = t \exp( \pm i \pi  \alpha y  ) $.  Noting
the invariance of the problem for translations along the x-direction we
introduce plane waves along the $ x $  axis and
define $ c_{m \hat x + n \hat y } = e^{i \nu m}  \tilde{c}_n  $ resulting
in a Hamiltonian for $  \tilde{c}_n $ of the form in 
\begin{equation} \label{eq:harponed}
 h_{1D} = - \sum_{r} ( \tilde{c}^{\dagger}_{r} \tilde{c}_{r+1} + cc ) - 
2 \sum_{r} \tilde{c}^{\dagger}_{r} \tilde{c}_{r} \cos{ ( 2 \pi r \alpha  - \nu  ) } .
\end{equation}
If we demand that this Hamiltonian be periodic with period $ q $
we find that that $ \alpha = p/q $ is the ratio of integers.  The flux per
square in the 2-D problem Eq. \ref{eq:harptwod} is thus a integral multiple of 
$ 2 \pi / q $.   Is this the transformation we have been doing
in the previous section?

If we begin with a $ p \times p $ periodic two-dimensional
lattice and add a constant magnetic field described in Landau gauge, we
find that $ \nu  $ must obey $ \nu = 2 \pi m /p $ with  $ m $ integer
and we obtain an equivalent set of $ p $ one dimensional problems Eq. \ref{eq:harponed}
with periodicity $ p$ indexed by the phase $ \nu $. In a conventional
formulation of the Harper problem we therefore cannot reduce the flux
quantum to sum to a single quantum over the entire $ 2-D $ lattice.
Indeed, the flux configuration in the 2-D Landau gauge is quite
different from that shown in Fig. \ref{fig:vortexplots} which 
describes a single flux quantum spread over the lattice and lacks
translational invariance in both $ x $ and $ y $ direction. (Note
the top row vertical bonds is not uniform along $ x $.)
A careful analysis of the gauge choice in Fig. \ref{fig:vortexplots} or
another clever choice of gauge other than Landau might possibly be used to show 
that the FFT remapping of the one-dimensional problem of length $ p^2 $ is
in fact Harper's problem,  but I have not been able to do this.

More digressions on the Harper problem would stray further from the
main point of the communication which is the presentation of a 
decoupling scheme of the tight binding model that is in principle
more general than Harper's problem. Unfortunately I can find no other
notrivial example than the relation between the one and two dimensional 
Harper model, but with flux and periodicity reduced from the ordinary
formulation of that problem.

\section*{Further work}
These proceedings has provided an opportunity to publish work which
I find interesting but admittedly incomplete.  When performing the
recursion on the tight binding model only, each FFT recursion
decouples the spatial degrees of freedom into two coupled blocks.
My original hope was that these transformation might be the basis
of a renormalization transformation; if an interaction could be
found that was similarly decoupled, a novel way to solve an interacting
model would be found. Unfortunately this goal has not been fulfilled
but perhaps someone might be motivated to look at this problem again
with the observations presented here.

\section*{Acknowledgment}
I thank Nihat Berker for being my friend, mentor and collaborator
during my first three years as a graduate student at Harvard. His
boundless energy and enthusiasm helped make these years very happy
and productive.  I also thank him for putting the necessary friendly
pressure on me to write up this contribution for these proceedings.
I also thank the referee for making several pertinent and important
comments that clarified the relation to Harper's equation.

\end{document}